\documentclass[10pt]{iopart}
\usepackage[T1]{fontenc}
\usepackage[numbers,sort&compress]{natbib}
\usepackage{color}
\usepackage{amssymb}
\usepackage{graphicx}
\usepackage{parskip}

\newcommand{\ra}[1]{\renewcommand{\arraystretch}{#1}}

\begin{document}

\title[]{The role of solvent interfacial structural ordering in maintaining stable graphene
dispersions}
\author{Urvesh Patil$^{1}$ and Nuala M. Caffrey$^{2, 3}$}
\address{$^1$ School of Physics \& CRANN Institute, Trinity College Dublin, Dublin 2, Ireland}
\address{$^2$ School of Physics, University College Dublin, Dublin 4, Ireland}
\address{$^3$ Centre for Quantum Engineering, Science, and Technology, University College Dublin, Dublin 4, Ireland}
\ead{nuala.caffrey@ucd.ie}

\begin{abstract}

Liquid phase exfoliation (LPE) is the most promising method for the low-cost, scalable production of two-dimensional nanosheets from their bulk counterparts. 
Extensive exfoliation occurs in most solvents due to the huge amount of energy introduced by sonication or shear mixing. However, the subsequent dispersion is not always stable, with extensive reaggregation occurring in some solvents. 
Identifying the optimal solvent for a particular layered material is difficult and requires a fundamental understanding of the mechanism involved in maintaining a stable dispersion. 
Here, we use molecular dynamics calculations to show that when graphene is immersed in a solvent, distinct solvation layers are formed irrespective of the choice of solvent and their formation is energetically favourable for all considered solvents. 
However, energetic considerations such as these do not explain the experimental solvent-dependence of the dispersion concentration. 
Instead, we find that solvents with high diffusion coefficients parallel to the graphene layer result in the lowest experimental concentration of graphene in solution. 
This can be explained by the enhanced ease of reaggregation in these solvents.
Solvents with smaller diffusion coefficients result in higher experimental graphene concentrations as reaggregation is prevented. 
In the low diffusion limit, however, this relationship breaks down. We suggest that here the concentration of graphene in solution depends primarily on the separation efficiency of the initial exfoliation step. Based on this, we predict that the concentration of exfoliated graphene in solvents such as benzaldehyde and quinoline, which have low diffusion constants, can be increased dramatically by careful tuning of the experimental sonication parameters. 
\end{abstract}

\noindent{\it liquid phase exfoliation; graphene, molecular dynamics; diffusion\/}

\ioptwocol

\section{Introduction}

The liquid-phase exfoliation (LPE) of layered materials is a scalable top-down method to produce industrial quantities of monolayer or few-layer two-dimensional (2D) nanosheets at a reasonable cost~\cite{nicolosi2013liquid, coleman2012liquid, lin2017liquid, raza2021advances}. Solvent-dispersed 2D nanosheets can find applications as inks for printed flexible electronics~\cite{secor2015rapid, torrisi2014electrifying}, energy generation and storage~\cite{worsley2018all}, sensing~\cite{boland2016sensitive} and fire resistant coatings~\cite{davesne2019hexagonal}.
The ability of LPE to produce large quantities compensates for relatively small lateral size flakes and low yield achieved compared to other exfoliation methods~\cite{li2020mechanisms}. However, the quality and quantity of the exfoliated layers depends dramatically on the solvent used~\cite{backes2017guidelines, hernandez2008high}. A careful choice of both the solvent and the sonication parameters can improve the monolayer yield and increase the lateral dimensions achieved, but a fundamental lack of understanding of the mechanism driving the exfoliation and stabilization processes impedes this optimisation procedure~\cite{tung2016graphene, khan2012size}.

These two processes, exfoliation and stabilization, can be considered separately even though they occur effectively simultaneously experimentally.
Exfoliation occurs when the van der Waals interactions binding the layers together are overcome via large shear forces that are introduced in the presence of a solvent through either sonication, high-shear mixing or wet-ball milling~\cite{li2020mechanisms}. The solvent has been suggested to play a relatively minor role in this initial separation of layers, with extensive exfoliation occurring in most solvents due to the huge amounts of energy involved~\cite{backes2019equipartition, xiang2023ultrasound}.
The solvent then stabilizes the exfoliated nanosheets preventing reaggregation and sedimentation. The choice of solvent now plays a critical role. 
Solvent exchange methods, where different solvents are used in the exfoliation and stabilization steps, find that those solvents which do not perform well in the initial exfoliation step may perform very well in maintaining a stable dispersion of already exfoliated monolayers~\cite{zhang2010dispersion, liang2010highly, li2012simple}. Evidently, the solvent plays a different role in the two steps but as they occur effectively simultaneously, their exact role is difficult to disentangle.

The majority of investigations in this regard concern the exfoliation of graphite. Hernandez et al.~compared the performance of a wide selection of solvents and found that the maximum graphene concentration was achieved by exfoliation in cyclopentanone, resulting in 8.5 $\pm$ 1.2 $\mu$g/ml~\cite{hernandez2009measurement}. Exfoliation in non-polar solvents, including toluene, heptane, hexane and  pentane, resulted in much lower graphene concentrations of 0.8 $\pm$ 0.4 $\mu$g/ml, 0.3 $\pm$ 0.4 $\mu$g/ml, 0.2 $\pm$ 0.1 $\mu$g/ml and 0.16 $\pm$ 0.05 $\mu$g/ml, respectively.
However, some highly polar molecules such as water and formamide also perform poorly, with the best performing solvents having a slightly polar nature (although not all slightly polar solvents perform well). 
The effectiveness of LPE for other layered materials, such as WS$_2$, MoS$_2$ and h-BN, is also strongly dependent on the choice of solvent~\cite{coleman2012liquid, coleman2011two}. 

Clearly, the inability to identify the optimal solvent for the exfoliation of a particular material without resorting to expensive trial-and-error is a limitation for its extensive industrial use. Understanding the nature of the interactions between the layered material and the solvent would allow for the preemptive screening of effective solvents for the exfoliation of layered materials. As well as aiming to maximize monolayer yield, there are other restrictions on the choice of solvent. For example, many of the better solvents for graphite exfoliation, including NMP, DMA and DMF, are toxic and facing restrictions by the European Chemicals Agency, rendering them unsustainable for future industrial use~\cite{echa}. Devising a screening tool to identify non-toxic, low-boiling point solvents which do not comprise on performance would be extremely useful~\cite{morton2023eco, salavagione2017identification}. Ideally these screening tools would be generalizable to
other two-dimensional materials beyond graphene.

One of the first attempts to screen for effective solvents found that the concentration of dispersed graphene was maximized for those solvents with a surface energy or tension similar to that of graphite, following the traditional surface wettability argument~\cite{hernandez2008high}. Although this method has had some successes, there are also examples where it breaks down~\cite{hernandez2009measurement}.
Hansen solubility parameters, characterized by three intermolecular interactions, namely hydrogen bonding, polar and nonpolar (dispersive), have also been used as screening parameters. Graphene was found to have Hansen dispersion, hydrogen bonding and polar solubility parameters of $\delta_D$ = 18 MPa$^{1/2}$ , $\delta_P$ = 9.3 MPa$^{1/2}$ , and $\delta_H$ = 7.7 MPa$^{1/2}$, respectively~\cite{hernandez2009measurement}. 
Searching for solvents with solubility parameters close to those of graphene led to the prediction and confirmation of cyclopenanone and cyclohexanone as excellent solvents of graphene. However, here again, there are some examples where
this screening technique fails -- dimethyl phthalate has very similar Hansen parameters to cyclopentanone, yet is a poor graphene solvent~\cite{hernandez2009measurement}. 
A peculiarity of the Hansen solubility parameters attributed to graphene is the non-zero polar component ($\delta_P = 9.3$ MPa$^{1/2}$). The mechanism which could lead to this is so-far unknown.
Contribution from edge-sites or the unintentional chemical functionalisation of the basal plane were touted as a potential physical mechanisms, yet there is little experimental evidence for either~\cite{backes2016spectroscopic, tao2017scalable}.

All of the above suggests that empirical solubility models, which consider only macroscopic solution thermodynamics, are blunt tools, missing some information about the stabilization mechanism~\cite{liang2018prediction, shen2015liquid}. These methods assume that kinetic effects driving  reaggregation play a minor role. 
There is growing acceptance that this is unlikely to be true and that explicit structural and electronic interactions between the solvent molecules and the solute are important in the stabilization of the exfoliated monolayers~\cite{goldie2022identification, lin2011molecular, mukhopadhyay2017ordering, mukhopadhyay2017deciphering}.

Here, we systematically study the origin of the solvent dependence of the stabilization of exfoliated monolayers.
We address several open questions: Is it sufficient to consider only energetic effects via
the enthalpy of mixing when choosing screening descriptors, what is the physical
mechanism leading to the experimental finding of a non-zero polar component of the
Hansen solubility parameter of graphene and, and how important are kinetic effects, such
as reaggregation, to the stabilization of graphene in a particular solvent?

\section{Methods}

\subsection{Choice of solvents}
The twelve solvents included in this investigation are listed in Table~\ref{tab:freeenergy}.
They were chosen to include those reported by Hernandez et~al.~\cite{hernandez2010measurement} to result in a wide range of graphene concentrations after exfoliation and include cyclopentanone (best performing solvent, polar aprotic), N-methyl-pyrrolidone (commonly used effective solvent, polar aprotic), benzaldehyde (poor performance, polar aprotic), bromobenzene (intermediate performance, slightly non-polar) and toluene (poor performance, non-polar). All others have intermediate exfoliation capabilities or polar properties. Note that while the concentration of exfoliated graphene in individual solvents has been increased considerably since the initial work of Hernandez et al, it is, to the best of our knowledge, the only available comparative study of the effectiveness of a wide range of solvents for graphite exfoliation.

\subsection{Molecular Dynamics}

Molecular dynamics calculations are performed using the Large-scale Atomic / Molecular Massively Parallel Simulator (LAMMPS) \cite{plimpton1995fast}. As the interaction between graphene and isolated solvent molecules was found to be primarily attractive dispersion forces~\cite{lazar2013adsorption,patil2018adsorption}, it can be modelled accurately using classical van der Waals (vdW) force fields.
The graphene monolayer is modeled as uncharged vdW spheres and are described by interaction parameters originally reported by Steele et al.~\cite{cheng1990computer} and later used to describe the interaction between graphene and molecular adsorbants~\cite{hardy2018design,lin2011molecular,fu2013molecular,kamath2012silico}. The molecules are described by the All Atom Optimized Potentials for Liquid Simulations (OPLS-AA) potentials, which were obtained from LigPargen web server~\cite{jorgensen2005potential,dodda20171,dodda2017ligpargen}.
The pair interaction coefficients between graphene and the molecules are obtained using Lorentz-Berthelot rules~\cite{lorentz1881ueber,berthelot1898melange}. Binding energies calculated using these potentials agree with experiment for small organic molecules adsorbed on graphene~\cite{lazar2013adsorption}.
The initial solvent structure was created using packmol~\cite{martinez2009packmol} in a box size of 105.91~\AA\ $\times$ 106.65~\AA\ $\times$ 120.0~\AA\ to reproduce the experimental room-temperature densities (see Table \ref{tab:data}). The graphene monolayer was then included in the simulation box by increasing the box size in the direction normal to the plane by (2 $\times$ 3.5)~\AA. The python package pymatgen was used to generate the LAMMPS structural data file~\cite{ong2013python}.
The systems were annealed from 600~K to 300~K for 50~fs with an integration timestep of 0.01~fs and then equilibriated at 300~K for 1~ns with an integration timestep of 1~fs. The system was further simulated with the NVE ensemble for 2~ns, of which 2000 samples from the last 1~ns were used to generate statistical distribution functions. For the duration of the complete simulation the graphene layer is held fixed.

\subsection{Thermodynamics of Mixing}

The Helmholtz free energy of mixing, $\Delta A_{\mathrm{mix}}$, is an indication of the favorable formation of a solvation structure. 
As $\Delta A_{\mathrm{mix}} = \Delta H_{\mathrm{mix}} - T \Delta S_{\mathrm{mix}}$, where $\Delta H_{\mathrm{mix}}$ and $\Delta S_{\mathrm{mix}}$ are the enthalpy and entropy of mixing, respectively, a requirement of mixing is that $\Delta H_{\mathrm{mix}} < T \Delta S_{\mathrm{mix}}$. Note that the PV term of the enthalpy is negligible for liquids and solids so it can be neglected. In this case, the internal energy of mixing at 0~K will be equal to the enthalpy of mixing at 0~K. 
$\Delta A_{\mathrm{mix}}$ was determined from molecular dynamics simulations using achemical free energy methods~\cite{duarte2017approaches}. 
Such methods simulate a series of non-physical intermediates to calculate the free energy of transferring a solute (here graphene) from vacuum (state 0) to solution (state 1). 
A continuous variable $\lambda$ parameterizes the path between state $0$ and state $1$. $\Delta A_{\mathrm{mix}}$ is calculated here using the finite difference thermodynamic integration (FDTI) method~\cite{mezei1987finite, kirkwood1935statistical, jorgensen1985monte}. MD simulations are carried out for a discrete set of $\lambda$ values and the free energy of mixing is calculated as:
$$\Delta A = \int_{\lambda = 0}^{\lambda = 1}\biggl< \frac{\partial U}{\partial \lambda}\biggr>_\lambda d\lambda \approx \sum_{i=0}^{N-1} w_i \biggl< \frac{U(\lambda_i + \delta) - U(\lambda_i)}{\delta} \biggr>$$
where $\langle \dots \rangle$ indicates an ensemble average, $w_i$ are integral weights and $\delta$ is the finite
difference parameter which satisfies $(\lambda_i - \lambda_{i-1}) >> \delta$. 
To avoid numerical instabilities the Leonard-Jones potential is modified to be a function of $\lambda$.
The system is simulated for 10~ns in the NVT ensemble at 300~K. The parameter $\lambda$ is incremented in 40 steps (every 250~ps) with the finite difference parameter $\delta = 0.001$. For each $\lambda$, the first 100~ps are discarded for equilibration and the ensemble average is calculated by sampling every 20~fs from the last 150~ps.
The enthalpy and entropy contributions to the total free energy are extracted from the slope and $y$-intercept of $\Delta A / T$ versus $1/T$, respectively, where the free energy is calculated at 290~K, 300~K and 310~K.

\subsection{Diffusion Coefficients}

The diffusion coefficient of solvent molecules confined in a region $[a, b]$ parallel to the
graphene sheet, $D_{xx}$, was determined from MD trajectories using the method described
in Liu et al.~\cite{liu2004calculation} and Agosta et al.~\cite{agosta2017diffusion}.
$D_{xx}$ is calculated as: 
$$
D_{xx}(\{a,b\}) = \lim_{\tau \rightarrow \infty} \frac{\langle \Delta x(\tau)^2 \rangle_{\{a, b\}}}{2\,\tau\,P(\tau)}
$$
where the terms on the right hand side can all be calculated using standard molecular dynamics calculations. $\langle \Delta x(\tau)^2 \rangle_{\{a, b\}}$ is the mean square displacement of particles in the direction parallel to the surface in a region $\{a, b\}$ in a time window $\tau$ and $P(\tau)$ is the probability that a particle remains in the region $\{a, b\}$ during that time, i.e, the survival probability. 
$\langle \Delta x(\tau)^2 \rangle_{\{a, b\}}$ is extracted from MD trajectories via:
$$
\langle \Delta x(\tau)^2 \rangle_{a, b} = \frac{1}{T} \sum_{t=1}^{T} \frac{1}{N(t)} \sum_{i \in \zeta(t, t+\tau)} (x_i(t+\tau) - x_i(t))^2
$$
where $\zeta(t, t+\tau)$ is the set of all particles that stay in the volume $a < z < b$ from time $t$ to $t + \tau$, $N(t)$ is the number of particles in the region $\{a,b\}$ at time $t$, and $T$ is the number of time steps to average. The survival probability is calculated as:
$$
P(\tau) = \frac{1}{T} \sum_t^T \frac{N(t, t+\tau)}{N(t)}
$$
where $N(t, t + \tau )$ is the number of particles that stay within the region $\{a,b\}$
between times $t$ and $t + \tau$ . Note that the parameters in the calculation of diffusion coefficient are $\tau$, $T$, and the definition of the region between $a$ and $b$.
Here, the equilibrated structure was simulated with the NVE ensemble
for $T=150$~ps and the center of mass of each molecule recorded at each time step. A time
window ($\tau$) of 35~ps is used to calculate the diffusion coefficient. $a$ and $b$  are chosen to encompass the entire first solvation shell of each system. 

\section{Results}

In Section~\ref{freeenergy}, we show that the free energy of mixing is always negative regardless of the solvent due to a large enthalphic contribution for solvating graphene. A significant entropic penalty, however, suggests that the solvent must undergo significant ordering in the presence of graphene. Section~\ref{solvationshell} discusses this structural and orientational ordering. Finally, Section~\ref{reaggregation} presents the diffusion coefficients of solvent molecules in the first solvation shell and discusses how these molecules play a major role in preventing the reaggregation of graphene sheets. 

\subsection{Free Energy of Mixing}\label{freeenergy}

It was previously shown, using \emph{ab initio} calculations, that the interaction between an isolated gas phase solvent molecule and a pristine graphene monolayer is van der Waals in nature, with no charge transfer involved. This is independent of the type of molecule, and the binding strength of a solvent molecule on graphene is not correlated with its ability to exfoliate graphite~\cite{patil2018adsorption}. 
To determine if instead the collective behaviour of the molecules near the surface of a graphene monolayer is the determining quantity, we calculate the Helmholtz free energy of mixing, $\Delta A_{\mathrm{mix}}$, of graphene in a solvent. If $\Delta A_{\mathrm{mix}} < 0$, then the mixture is thermodynamically more stable than each of the two components separately.  
The Helmholtz free energies of mixing for graphene at room temperature are shown in the first column of Table \ref{tab:freeenergy} for a variety of solvents.  
\begin{table*}[t!]\centering
	\ra{1.2}
	\setlength{\tabcolsep}{6pt} % Default value: 6pt
	\begin{tabular}{@{\extracolsep{3pt}}lccc@{}}
		\hline \hline 
		Solvent &   $\Delta A_{\mathrm{mix}}$ (MJ/mol) & $\Delta H_{\mathrm{mix}}$ (MJ/mol) & $T \Delta S_{\mathrm{mix}}$ (MJ/mol) \\ \hline \hline 
		Quinoline & -11.08 & -28.32 & -17.13 \\	
		Bromobenzene & -10.41 & -9.57 & 0.83 \\
		NMP & -10.172 & -28.15 & -18.01 \\
		Benzaldehyde & -9.90 & -26.57 & -16.64 \\
		Chlorobenzene & -9.65 & -23.10 & -13.45 \\  
		1,3-dioxolane & -9.22 & -25.23 & -16.0 \\
		Toluene & -8.99 & -21.10 & -12.07 \\
		Benzonitrile & -8.94 & -21.98 & -13.06 \\
		Cyclopentanone & -8.33 & -24.07 & -15.74 \\
		Cyclohexanone & -7.57 & -22.26 & -14.69 \\
		\hline
	\end{tabular}
	\caption{\label{tab:freeenergy} The Helmholtz free energy of mixing, $\Delta A_{\mathrm{mix}}$, the corresponding enthalpy, $\Delta H_{\mathrm{mix}}$ and the entropy term calculated at room temperature (300~K), $T \Delta S_{\mathrm{mix}}$. These values are determined for a graphene area of $112.953$~nm$^2$.}
\end{table*}
At 300~K, $\Delta A_{\mathrm{mix}}$ is always negative, irrespective of the solvent, suggesting the favorable mixing of monolayer graphene in each of the solvents. 
It is most favourable in quinoline ($\Delta A_{\mathrm{mix}} = -11.08$~MJ/mol) followed by bromobenzene ($\Delta A_{\mathrm{mix}} = -11.41$~MJ/mol). It is least favourable in cyclohexanone, at $-7.57$~MJ/mol.
The magnitudes of $\Delta A_{\mathrm{mix}}$ reported here are consistent with those reported by Oyer et al.~for the free energy of mixing of graphene in benzene and hexaflourene~\cite{oyer2012stabilization} and by Mukhopadhyay et al.~for the free energy of mixing of graphene in water and DMSO~\cite{mukhopadhyay2019exfoliation}.

The decomposition of the Helmholtz energy into the enthalpic and entropic contributions are given in the second and third column of Table \ref{tab:freeenergy}, respectively. 
In almost all cases there is a significant entropic penalty for solvating graphene, with $T \Delta S_{\mathrm{mix}}$ lying between $-17.13$~MJ/mol and $-12.07$~MJ/mol.
This entropic penalty is compensated by a large enthalpy change ranging between $-28.32$~MJ/mol in quinoline to $-21.10$~MJ/mol in toluene, driven by strong van der Waals interactions. Bromobenzene is an exception to this: $\Delta H_{mix}$ is much smaller in magnitude than for the other solvents at $-9.57$~MJ/mol, while the entropy contribution is positive at 0.83~MJ/mol. 
However, this unusual behavior is not evident in the overall Helmoltz free energy of mixing which is similar to that of the other considered solvents, at $-10.41$~MJ/mol. 

In general, no correlation can be found between the magnitude of $\Delta A_{\mathrm{mix}}$ and the experimentally-determined graphene concentrations after exfoliation. 
As such, the Helmoltz free energy of mixing cannot be used as a mechanism of screening for more effective solvents.
Instead, the calculated free energies indicate that the solvent-dependent exfoliation is not energetically bound and other dynamical effects should be considered.

\subsection{Solvation Shell Formation}\label{solvationshell}

\begin{figure*}
	\centering
	\includegraphics[width = 0.9\textwidth]{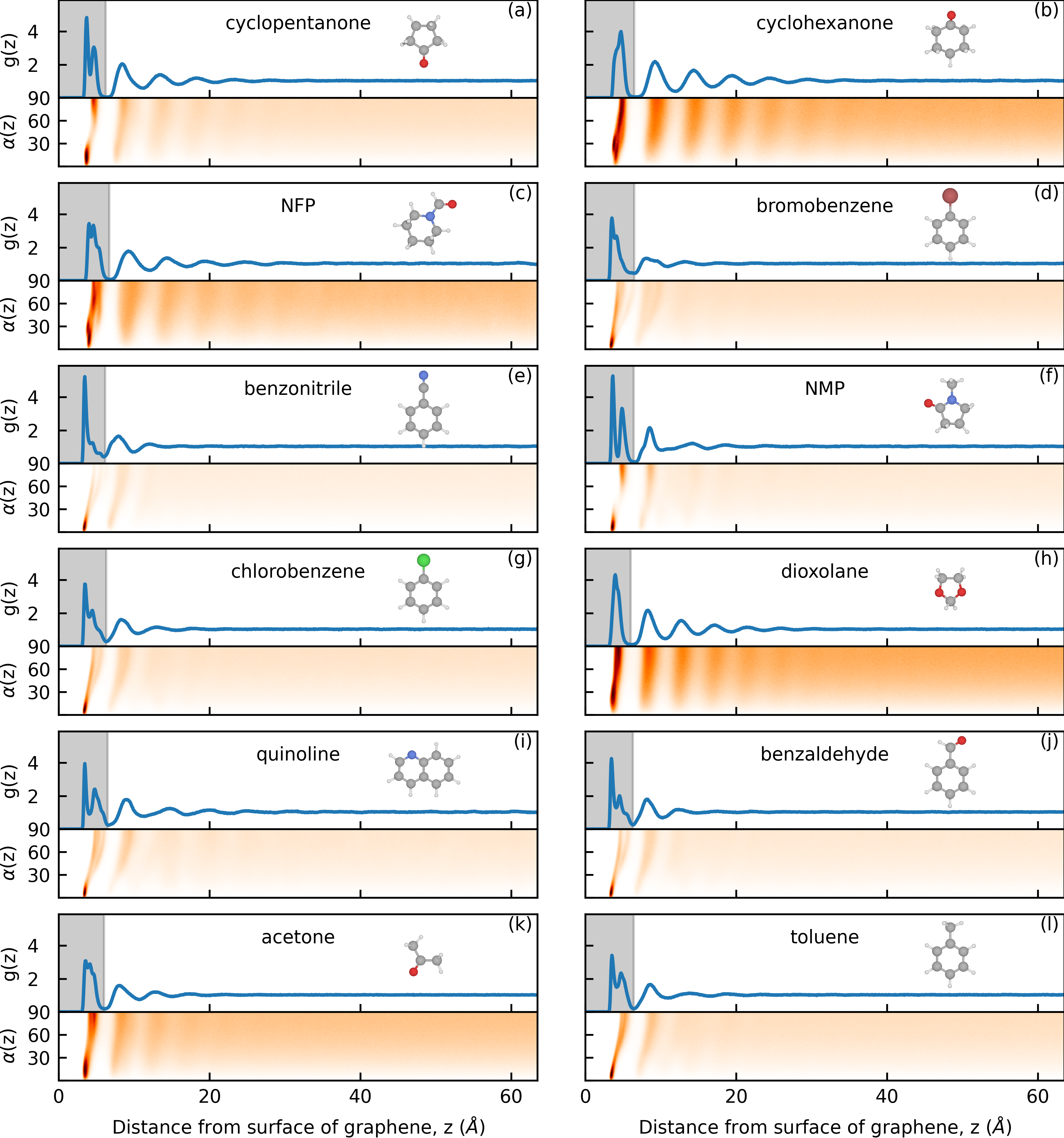}
	\caption{Pair distribution functions. (Top panel) $g(z)$ is the position distribution of the center of mass of the solvent molecules. $g(z)$ is given relative to the bulk solvent number density. (Bottom panel) $\alpha (z)$ is the probability distribution function of the angle made by the normal to the plane of molecule with the axis perpendicular to the plane of graphene. The graphene monolayer is located at $z = 0$~\AA.}
	\label{fig:plane}
\end{figure*}

The entropic penalty suggests that there must be some solvent reconstruction or ordering that occurs due to the presence of graphene. 
To understand this, the location and orientation of solvent molecules close to the graphene layer are extracted from molecular dynamics calculations. 
The position distribution function $g(z)$ and the angle distribution function $\alpha (z)$ are shown in Fig.~\ref{fig:plane}. 
The position distribution function $g(z)$ is given relative to the bulk solvent number density so that a value greater than 1 indicates an accumulation of molecules relative to the bulk solvent, and a value less than 1 indicates a depletion of molecules relative to the bulk.
$\alpha(z)$ is the angle made by the normal to the plane of the molecule with respect to the $z$ axis, where the plane is defined as containing at least 3 atoms of the molecular ring system.

Irrespective of the polar nature of the molecule, distinct solvation shells are formed next to the graphene layer as a result of surface confinement, molecule - molecule interactions and molecule - graphene interactions. The first solvation shell (grey shaded area) can be recognized as a peak in $g(z)$ followed by a deep trough near the surface of graphene. 
The peak associated with the first solvation shell is also the sharpest, indicating that the highest degree of transverse ordering occurs for those closest to graphene.
The first solvation shell has a complex structure in most cases due to the adsorption geometry of the molecules on the graphene surface. 
This is particularly evident for cyclopentanone (Fig~\ref{fig:plane}(a)) and NMP (Fig~\ref{fig:plane}(f)) where the first solvation shell is composed of two distinct peaks. 
The reason for this can be found by looking at $\alpha (z)$: the solvent adsorbs on graphene with two distinct angular orientations -- one with the molecular plane lying approximately parallel to the surface ($\alpha = 0^\circ$) and one with the plane almost perpendicular ($\alpha = 90^\circ$).
As the distribution function is based on the centre of mass of the molecules, this manifests as a double peak structure in $g(z)$. As a result, the dipole moments of the molecules also show two distinct polar angles $(\Theta (z))$, whereas the azimuthal angle $(\phi (z))$ shows no distinct orientation preference indicating no in-plane ordering of the dipoles (see Fig~\ref{fig:rdf}). 
This behavior was previously found in molecular dynamic calculations involving NMP interacting with both graphene and carbon nanotubes  \cite{frolov2012molecular,fu2013molecular,terrones2016enhanced}.
In other cases, the distinction between the two peaks is not as clear. 
For example, the first solvation shell of cyclohexanone (Fig~\ref{fig:plane}(b)) has a small peak corresponding to those molecules located closest to graphene and orientated parallel to it, but the remainder of the molecules in the first solvation shell do not exhibit an angular preference. 
This is also the case for NFP, bromobenzene, chlorobenzene, dioxolane, quinoline, acetone and toluene. 
For the case of benzonitrile and benzaldehyde the first solvation shell is comprised almost entirely of molecules lying parallel to the graphene layer, with no second peak visible. 
There is some experimental evidence for the confinement of solvent molecules near the surface of graphene - Arunachalam
et al. showed that the NMP molecules near the surface of graphene show a reduction
in rotational degrees of freedom using rotating frame Overhauser effect spectroscopy –
nuclear magnetic resonance technique~\cite{arunachalam2018graphene}. 

\begin{figure}
	\centering
	\includegraphics[width= \columnwidth]{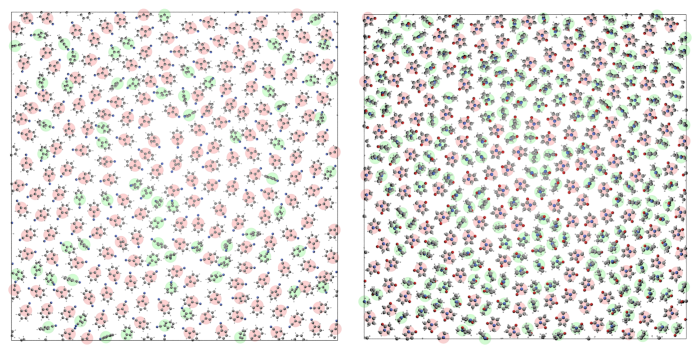}
	\caption{A snapshot of molecules in first solvation shell of benzonitrile (left) and NMP (right) taken from the molecular dynamics simulation. Red shading indicates molecules belonging to the first peak in the solvation shell and green shading indicates the rest of the molecules.}
	\label{fig:sidemd}
\end{figure}

To visualize this orientational ordering, Fig~\ref{fig:sidemd} shows a snapshot from the molecular dynamics simulation showing only the first solvation shell of benzonitrile and NMP. All molecules belonging to the very first peak are shaded in red, and all others are shaded in green. As expected in the case of benzonitrile, it can be seen that essentially all molecules in the first peak are orientated parallel to the plane of graphene $(\alpha \approx 0^\circ)$ with very few molecules adsorbed perpendicularly. In contrast, a small but significant number of NMP molecules are adsorbed perpendicular to the sheet. These molecules appear to fill all the gaps that exist in between the flat-lying molecules. 

\begin{figure}[t]
	\centering
	\includegraphics[width = 0.95\columnwidth]{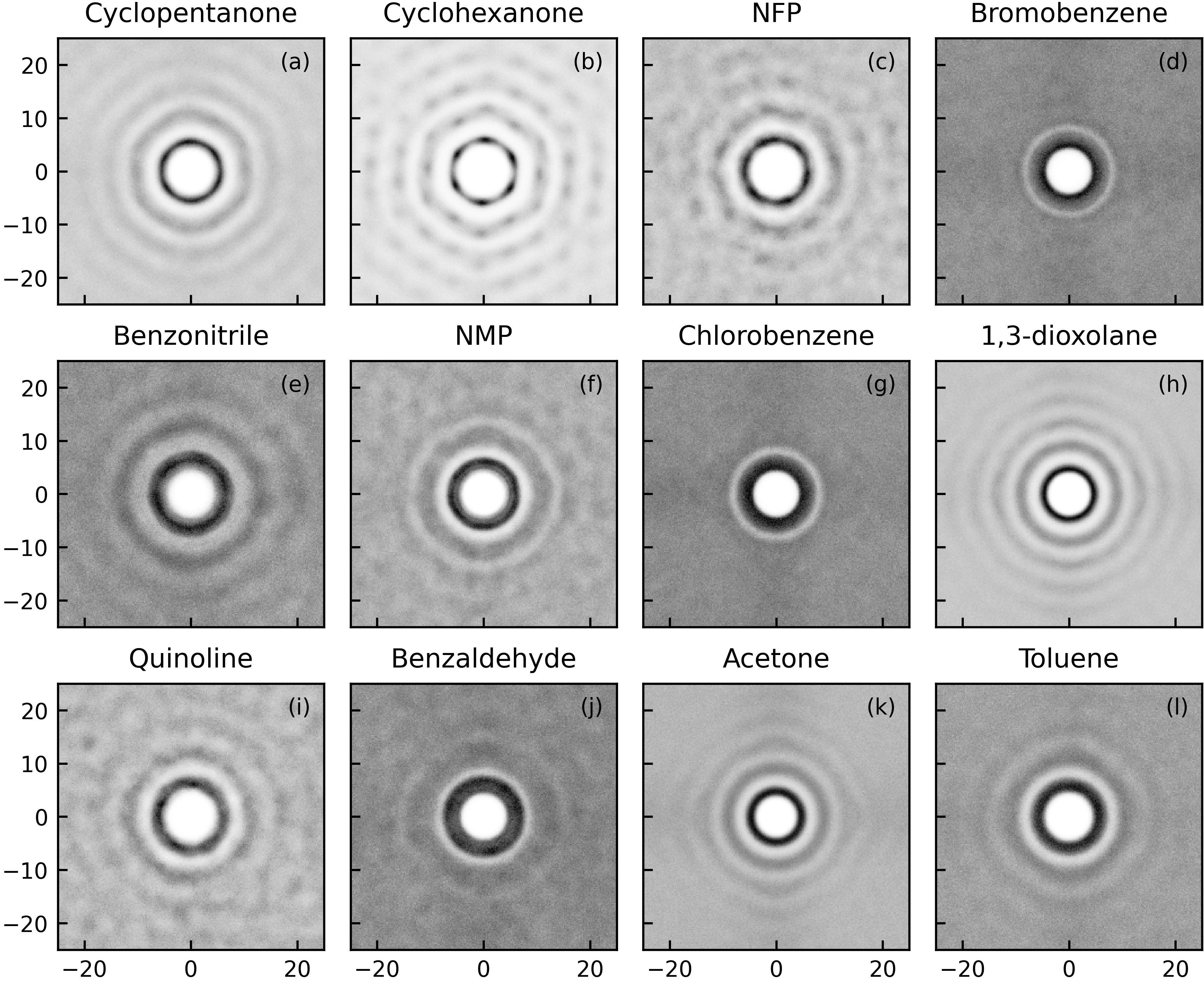}
	\caption{Pair correlation function for the center of geometry of the solvent molecules present in the first solvation shell.}
	\label{fig:center}
\end{figure}

To determine if any long-range lateral order is present, we calculated the pair correlation functions for the geometric centers for the molecules in the first solvation shell. These are shown in Fig~\ref{fig:center}. In all cases, there is a local ordering around the center (0,0) due to the presence of an exclusion zone around the molecule. This ordering then decays rapidly moving away from the center. The strongest ordering is seen in case of cyclopentanone, cyclohexanone and 1,3-dioxolane. Intermediate ordering behaviour is seen in case of NFP, benzonitrile, NMP, quinoline, acetone and toluene, with no ordering beyond the first shell seen in case of bromobenzene, chlorobenzene and benzaldehyde. 
This is in agreement with the work of Terrones et al.~who found that while a monolayer of NMP molecules confined between two graphene sheets exhibit long-range hexagonal ordering, this ordering is lost as the thickness of the molecular layer increases \cite{terrones2016enhanced}.

In response to the formation of the first molecular layer, the rest of the solvent then reorganizes to form further solvation shells, as seen in Fig~\ref{fig:plane}. 
As expected, the amplitude of the solvation shell peaks decay as the influence of the solute wanes, finally approaching a constant value of 1, indicating bulk solvent behaviour. 
The distance between graphene and the nearest solvent atoms ($\mathrm{h}_{\mathrm{min}}$), the number of solvation shells $(\mathrm{N}_{\mathrm{shells}})$, the depth to which the solvation shells extend away from graphene, and the average distance between two consecutive solvation shells are given in Table~\ref{tab:mddata}. 
As the distance between graphene and the atom of the molecule nearest it is determined by the vdW interaction, it is very similar across all solvents regardless of the molecular orientations, varying from 1.99~\AA\ for bromobenzene to 2.07~\AA\ for benzaldehyde.
The number of solvation shells varies from a minimum of 3 in the cases of bromobenzene, benzonitrile, chlorobenzene and benzaldehyde to a maximum of 6 for cyclohexanone and 1,3-dioxolane, where in counting the total number of solvation shells we have defined the last shell as the final peak that deviates from bulk by at most 5\%.  The average distance between adjacent solvation shells ranges from 4.25 \AA\ for benzonitrile to 5.25 \AA\ for quinoline. 

\begin{table}\centering
	\ra{1.2}
	\setlength{\tabcolsep}{6pt} % Default value: 6pt
	\begin{tabular}{@{\extracolsep{3pt}}lcccc@{}}
		\hline \hline 
		Molecules & h$_{\mathrm{min}}$& N$_{\mathrm{shells}}$ & Depth (\AA) & Period (\AA) \\ \hline \hline 
		
		Cyclopentanone & 2.03 & 5 & 23.34 & 4.97  \\
		Cyclohexanone & 2.02 & 6 & 29.31 & 5.02  \\
		NFP & 2.0 & 5 & 24.67 & 5.14  \\
		Bromobenzene & 1.99 & 3 & 13.11 & 4.95  \\
		Benzonitrile & 2.03 & 3 & 12.16 & 4.25  \\
		NMP & 2.0 & 4 & 18.7 & 5.08  \\
		Chlorobenzene & 2.02 & 3 & 12.92 & 4.64  \\
		1,3-dioxolane & 2.05 & 6 & 25.94 & 4.41  \\
		Quinoline & 2.0 & 5 & 24.86 & 5.25  \\
		Benzaldehyde & 2.07 & 3 & 12.41 & 4.38  \\
		Acetone & 2.02 & 4 & 17.37 & 4.7  \\
		Toluene & 2.01 & 4 & 18.51 & 4.95  \\
		\hline
	\end{tabular}
	\caption{\label{tab:mddata} Data extracted from the pair distribution function $g(z)$ shown in Fig.~\ref{fig:plane}. h$_{\mathrm{min}}$ is the distance between graphene and the solvent atoms closest to it. N$_{\mathrm{shells}}$ indicates the number of solvation shells present where the final solvation shell is defined to that peak which deviates from the bulk value by at most 5\%. Depth is the distance of last solvation shell from graphene. Period indicates the average periodicity of the solvation shells.}
\end{table}

Despite these differences, it is clear from Fig~\ref{fig:plane} that the solvation structure cannot be used to explain the observed phenomenon of solvent-dependent graphene stabilization. 
For example, there is no appreciable difference between the solvation structure of NMP and toluene, one of the best and worst solvents for the exfoliation of graphite, respectively.
However, the existence of these solvation layers can explain the origin of the non-polar Hansen solubility parameter attributed to graphene. Rather than to graphene alone, the solubility parameters can be attributed to an effective solute comprised of graphene and its first few solvation shells. Due to partial charges on the atoms of the first solvation shell, the effective solute appears to have surface charges, resulting in a designation of a non-zero polar solubility parameter. 

\subsection{Molecular Diffusion}\label{reaggregation}

In the analysis up until now, we have assumed that the process of stabilization is thermostatically driven, and kinetic effects such as sedimentation play a minor role.
This may not be necessarily true, and in certain conditions may be the most important consideration.
If the solvent-dependence of graphene stability in a dispersion is determined by reaggregation effects, then screening descriptors based on energetic considerations alone will not be not appropriate. 
Long-term stability can only be obtained if the dispersion is kinetically stabilized against reaggregation when layer collision occurs. 
If the layers are far apart, this will be influenced by the viscosity of the solvent \cite{salavagione2017identification, gilliam2021evaluating}.
When the layers are close together, the diffusion coefficient of molecules confined close to the graphene sheet will determine the ease at they can be ejected to facilitate reaggregation. 
Indeed, the addition of surfactants or polymers to graphene dispersions to prevent reaggregation operate on this principle~\cite{griffin2020effect, hu2021dispersant}.

\begin{figure}
	\centering
 	\includegraphics[width= \columnwidth]{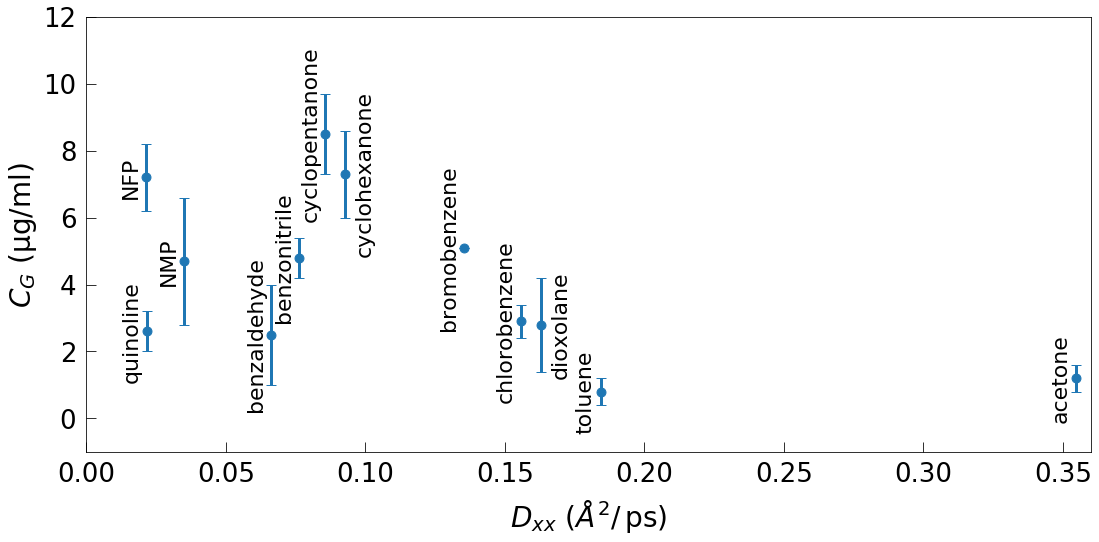}
	\caption{The diffusion coefficient of solvent molecules in the first solvation shell in the direction parallel to the surface of graphene as a function of the experimental concentration of graphene $C_G$ ($\mathrm{\mu}$g/ml$)$ in that solvent. Experimental data is taken from Herndandez et al.~\cite{hernandez2010measurement}.}
	\label{fig:diffusion}
\end{figure}

To determine the extent to which solvent molecules can kinetically block reaggregation, we calculate the diffusion coefficient in the direction parallel to the graphene surface of the molecules in the first solvation shell. 
The results are shown in Fig.~\ref{fig:diffusion} as a function of the experimental graphene concentrations taken from Ref.~\cite{hernandez2010measurement}.
The highest parallel diffusion coefficient is found for acetone, at 0.35 \AA$^2$/ps. This is followed by toluene (0.18 \AA$^2$/ps), dioxolane (0.16 \AA$^2$/ps) and chlorobenzene (0.16 \AA$^2$/ps). It is notable that all of these are poor solvents for the exfoliation of graphite, with the worst solvent corresponding to that with the largest parallel diffusion coefficient. As the diffusion constant decreases, the experimental concentration increases, with the best solvent (cyclopentanone) having a parallel diffusion constant of 0.085 \AA$^2$/ps.
However, in the very-low diffusion regime the concentration does not follow this upward trend. As the molecular hindrance to re-aggregation is higher in this regime, one would expect that the exfoliated layers would remain in the dispersion resulting in an increased monolayer concentration. 
Instead, the concentration decreases again. 

While this requires further investigation, we hypothesis that the concentration of graphene is now dependent on the ability of that solvent to faciliate the exfoliation process itself, as opposed to the stabiliziation process.  
When graphene is dispersed in a solvent with a low surface diffusion coefficient, such as benzaldehyde, the reaggregation rate will be low and the concentration of graphene layers will be maintained at its initial concentration. 
In contrast, when the diffusion coefficient is high, the graphene concentration will always be low due to a high rate of reaggregation. Such solvents will result in a low concentration dispersion even if a large number of monolayers are initially exfoliated due to efficient reaggregation. 
As a consequence then, solvents with low surface diffusion coefficients, such as quinoline and benzaldehyde, are good candidate for the optimization of the initial separation process. If the initial concentration of graphene in these solvents can be increased, the low diffusion rate in the first solvation shell means that reaggregation will be hindered and the high concentration maintained.

\section{Conclusion}
We find that there is a molecular-level structural and orientational ordering of solvent molecules at the graphene-solvent interface which extends up to 30~\AA\ away from the graphene layer. These solvation layers form in all cases and behaves similarly irrespective of the solvent polarity. 
We speculate that the non-polar Hansen parameter obtained by fitting the experimental data can be attributed to the formation of the first solvation shell as the molecules in the first solvation shell and the graphene form an effective solute.
The formation of these solvation layers is spontaneous as indicated by the free energy and its decomposition in enthalpy and entropy contributions. This independence of the solvation structure on the nature of the solvent molecule suggests that non energetic contributions are responsible for the observed solvent-dependence of graphene stabilization, and specifically the kinetics of reaggregation. We propose that the
diffusion coefficient of molecules in the first solvation shell is an appropriate property to determine the probability of reaggregation in a solvent.
We note that in the low-diffusion limit, this appears to break-down, at least for the solvents investigated here. This origin of this requires further work, but could be related to the role of the solvent in the initial separation step.
We predict that the concentration of graphene exfoliated in solvents with low diffusion coefficients parallel to the graphene plane, such as benzaldehyde and quinoline, can be enhanced dramatically by careful tuning of the experimental sonication parameters, as any produced monolayers will be prevented from reaggregation by steric effects due to the slowly diffusing solvent molecules on the monolayer surface.

\ack
This work was supported by a Science Foundation Ireland Starting Investigator Research Grant (15/SIRG/3314). Computational resources were provided by the supercomputer facilities at the Trinity Center for High Performance Computing (TCHPC) and at the Irish Center for High-End Computing (ICHEC).

\appendix

\section{Polar and azimuthal angle distribution functions}
\begin{figure*}[h]
	\centering
	\includegraphics[width = \textwidth]{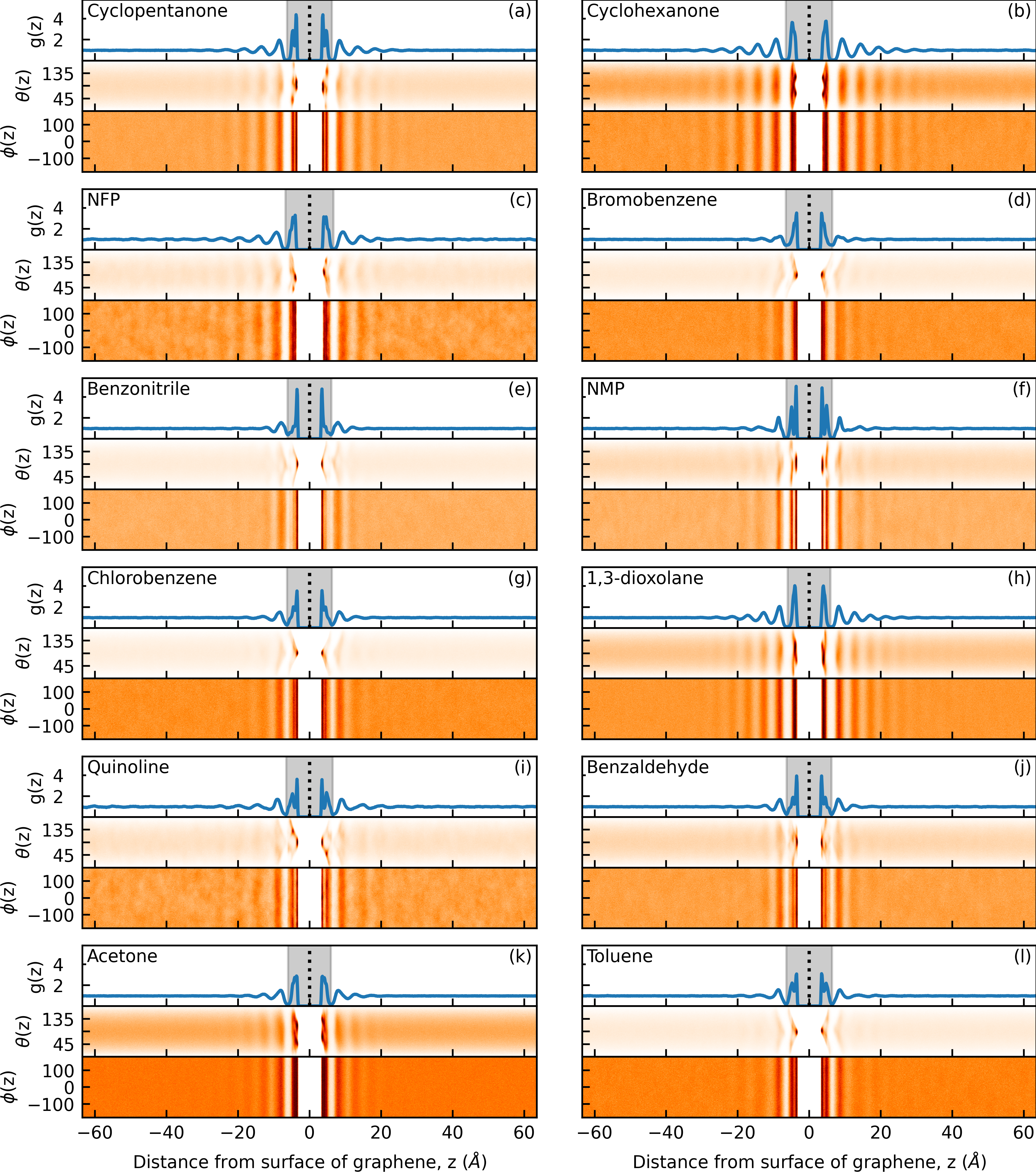}
	\caption{Pair distribution functions. (Top panel) $g(z)$ a histogram of the center of mass of the solvent molecules normalised by $(\frac{N}{L} dz)$ where $N$ is the number of molecules in the unit cell, $L$ is the length of the unit cell and $dz$ is the bin width. The graphene monolayer is located at $z = 0$~\AA. (This is the same as presented in Fig.~\ref{fig:plane}). $\theta (z)$ is the probability distribution function of the polar angle of the molecular dipole, where $90^\circ$ corresponding to the molecular dipole in the plane parallel to graphene and $0^\circ$ ($180^\circ$) corresponds to the molecular dipole in the plane perpendicular to the plane of graphene. (Bottom panel) $\phi (z)$ is the probability distribution function of the azimuthal angle of the molecular dipole, indicating the orientation of the dipole in the plane parallel to the plane of graphene, with $0^\circ$ ($180^\circ$) corresponding to the orientation along positive (negative) $x$-axis, $90^\circ$ ($-90^\circ$) corresponding to the orientation along the positive (negative) $y$-axis.}
	\label{fig:rdf}
\end{figure*}

The position distribution function $g(z)$, the polar angle distribution function $\theta(z)$ and the azimuthal angle distribution function $\phi(z)$ of the molecular dipoles close to graphene are shown in Fig\ref{fig:rdf}. The graphene layer is located at 0~\AA. The polar angle is measured with respect to the axis aligned to the positive direction (towards the right). $g(z)$ is given relative to the bulk solvent number density so that a value greater (less) than 1 indicates an accumulation (depletion) of molecules relative to the bulk solvent. $\theta(z)$ is the probability distribution function of the polar angle of the molecular dipole, where 90$^\circ$
corresponds to the molecular dipole in the plane parallel to graphene and 0$^\circ$ (or 180$^\circ$)
corresponds to the molecular dipole in the plane perpendicular to the plane of graphene.
$\phi(z)$ is the probability distribution function of the azimuthal angle of the molecular dipole, indicating the orientation of the dipole in the plane parallel to the plane of graphene, with 0$^\circ$ (or 180$^\circ$) corresponding to the orientation
along positive (negative) x-axis, 90$^\circ$ (-90$^\circ$) corresponding to the orientation along the positive (negative) y-axis. 

\section{Solvent Densities and Dipole Moments}
\begin{table}[h]\centering
	\ra{1.2}
	\setlength{\tabcolsep}{6pt} % Default value: 6pt
	\begin{tabular}{@{\extracolsep{3pt}}lcc@{}}
		\hline \hline 
		Molecules & density (g/ml) & dipole moment (D)  \\ \hline \hline 
		
		Cyclopentanone & 0.95 & 3.28 \\
		Cyclohexanone & 0.95 & 2.9 \\
		NFP & 1.02 & - \\
		Bromobenzene & 1.5 & 1.74 \\
		Benzonitrile & 1.01 & 3.2 \\
		NMP & 1.03 & 4.1 \\
		Chlorobenzene & 1.11 & 1.55 \\
		1,3-dioxolane & 1.06 & 1.19 \\
		Quinoline & 1.09 & 2.0 \\
		Benzaldehyde & 1.04 & 2.89 \\
		Acetone & 0.78 & 2.9 \\
		Toluene & 0.87 & 0.31 \\
		\hline
	\end{tabular}
	\caption{\label{tab:data} Experimental densities and dipole moments of the solvent molecules considered here extracted from the on-line databases~\cite{DDBONLINE,kim2019pubchem}.}
\end{table}

%\section*{References}
\bibliographystyle{iopart-num.bst}
\bibliography{bibliography}

\providecommand{\newblock}{}
\begin{thebibliography}{10}
\expandafter\ifx\csname url\endcsname\relax
  \def\url#1{{\tt #1}}\fi
\expandafter\ifx\csname urlprefix\endcsname\relax\def\urlprefix{URL }\fi
\providecommand{\eprint}[2][]{\url{#2}}

\bibitem{nicolosi2013liquid}
Nicolosi V, Chhowalla M, Kanatzidis M~G, Strano M~S and Coleman J~N 2013 {\em
  Science\/} {\bf 340}

\bibitem{coleman2012liquid}
Coleman J~N 2012 {\em Accounts of chemical research\/} {\bf 46} 14--22

\bibitem{lin2017liquid}
Lin S, Chui Y, Li Y and Lau S~P 2017 {\em FlatChem\/} {\bf 2} 15--37

\bibitem{raza2021advances}
Raza A, Hassan J~Z, Ikram M, Ali S, Farooq U, Khan Q and Maqbool M 2021 {\em
  Advanced Materials Interfaces\/} {\bf 8} 2002205

\bibitem{secor2015rapid}
Secor E~B, Ahn B~Y, Gao T~Z, Lewis J~A and Hersam M~C 2015 {\em Advanced
  Materials\/} {\bf 27} 6683--6688

\bibitem{torrisi2014electrifying}
Torrisi F and Coleman J~N 2014 {\em Nature nanotechnology\/} {\bf 9} 738--739

\bibitem{worsley2018all}
Worsley R, Pimpolari L, McManus D, Ge N, Ionescu R, Wittkopf J~A, Alieva A,
  Basso G, Macucci M, Iannaccone G {\em et~al.\/} 2018 {\em ACS nano\/} {\bf
  13} 54--60

\bibitem{boland2016sensitive}
Boland C~S, Khan U, Ryan G, Barwich S, Charifou R, Harvey A, Backes C, Li Z,
  Ferreira M~S, M{\"o}bius M~E {\em et~al.\/} 2016 {\em Science\/} {\bf 354}
  1257--1260

\bibitem{davesne2019hexagonal}
Davesne A~L, Lazar S, Bellayer S, Qin S, Grunlan J~C, Bourbigot S and Jimenez M
  2019 {\em ACS Applied Nano Materials\/} {\bf 2} 5450--5459

\bibitem{li2020mechanisms}
Li Z, Young R~J, Backes C, Zhao W, Zhang X, Zhukov A~A, Tillotson E, Conlan
  A~P, Ding F, Haigh S~J {\em et~al.\/} 2020 {\em ACS nano\/} {\bf 14}
  10976--10985

\bibitem{backes2017guidelines}
Backes C, Higgins T~M, Kelly A, Boland C, Harvey A, Hanlon D and Coleman J~N
  2017 {\em Chemistry of materials\/} {\bf 29} 243--255

\bibitem{hernandez2008high}
Hernandez Y, Nicolosi V, Lotya M, Blighe F~M, Sun Z, De S, McGovern I, Holland
  B, Byrne M, Gun'Ko Y~K {\em et~al.\/} 2008 {\em Nature nanotechnology\/} {\bf
  3} 563--568

\bibitem{tung2016graphene}
Tung T~T, Yoo J, Alotaibi F~K, Nine M~J, Karunagaran R, Krebsz M, Nguyen G~T,
  Tran D~N, Feller J~F and Losic D 2016 {\em ACS applied materials \&
  interfaces\/} {\bf 8} 16521--16532

\bibitem{khan2012size}
Khan U, O’Neill A, Porwal H, May P, Nawaz K and Coleman J~N 2012 {\em
  Carbon\/} {\bf 50} 470--475

\bibitem{backes2019equipartition}
Backes C, Campi D, Szydlowska B~M, Synnatschke K, Ojala E, Rashvand F, Harvey
  A, Griffin A, Sofer Z, Marzari N {\em et~al.\/} 2019 {\em ACS nano\/} {\bf
  13} 7050--7061

\bibitem{xiang2023ultrasound}
Xiang K, Huang S, Song H, Bazhenov V, Bellucci V, Birnsteinova S, de~Wijn R,
  Koliyadu J~C, Koua F~H, Round A {\em et~al.\/} 2023 {\em arXiv preprint
  arXiv:2305.08538\/}

\bibitem{zhang2010dispersion}
Zhang X, Coleman A~C, Katsonis N, Browne W~R, Van~Wees B~J and Feringa B~L 2010
  {\em Chemical Communications\/} {\bf 46} 7539--7541

\bibitem{liang2010highly}
Liang Y~T and Hersam M~C 2010 {\em Journal of the American Chemical Society\/}
  {\bf 132} 17661--17663

\bibitem{li2012simple}
Li J, Ye F, Vaziri S, Muhammed M, Lemme M~C and {\"O}stling M 2012 {\em
  Carbon\/} {\bf 50} 3113--3116

\bibitem{hernandez2009measurement}
Hernandez Y, Lotya M, Rickard D, Bergin S~D and Coleman J~N 2009 {\em
  Langmuir\/} {\bf 26} 3208--3213

\bibitem{coleman2011two}
Coleman J~N, Lotya M, O’Neill A, Bergin S~D, King P~J, Khan U, Young K,
  Gaucher A, De S, Smith R~J {\em et~al.\/} 2011 {\em Science\/} {\bf 331}
  568--571

\bibitem{echa}
ECHA 2020 Candidate list of substances of very high concern for authorisation
  \url{https://echa.europa.eu/candidate-list-table}

\bibitem{morton2023eco}
Morton J~A, Kaur A, Khavari M, Tyurnina A~V, Priyadarshi A, Eskin D~G, Mi J,
  Porfyrakis K, Prentice P and Tzanakis I 2023 {\em Carbon\/} {\bf 204}
  434--446

\bibitem{salavagione2017identification}
Salavagione H~J, Sherwood J, Budarin V, Ellis G, Clark J, Shuttleworth P {\em
  et~al.\/} 2017 {\em Green Chemistry\/} {\bf 19} 2550--2560

\bibitem{backes2016spectroscopic}
Backes C, Paton K~R, Hanlon D, Yuan S, Katsnelson M~I, Houston J, Smith R~J,
  McCloskey D, Donegan J~F and Coleman J~N 2016 {\em Nanoscale\/} {\bf 8}
  4311--4323

\bibitem{tao2017scalable}
Tao H, Zhang Y, Gao Y, Sun Z, Yan C and Texter J 2017 {\em Physical Chemistry
  Chemical Physics\/} {\bf 19} 921--960

\bibitem{liang2018prediction}
Liang K~Y and Yang W~D 2018 {\em AIP Advances\/} {\bf 8} 015018

\bibitem{shen2015liquid}
Shen J, He Y, Wu J, Gao C, Keyshar K, Zhang X, Yang Y, Ye M, Vajtai R, Lou J
  {\em et~al.\/} 2015 {\em Nano letters\/} {\bf 15} 5449--5454

\bibitem{goldie2022identification}
Goldie S~J, Degiacomi M~T, Jiang S, Clark S~J, Erastova V and Coleman K~S 2022
  {\em ACS nano\/} {\bf 16} 16109--16117

\bibitem{lin2011molecular}
Lin S, Shih C~J, Strano M~S and Blankschtein D 2011 {\em Journal of the
  American Chemical Society\/} {\bf 133} 12810--12823

\bibitem{mukhopadhyay2017ordering}
Mukhopadhyay T~K and Datta A 2017 {\em The Journal of Physical Chemistry C\/}
  {\bf 121} 10210--10223

\bibitem{mukhopadhyay2017deciphering}
Mukhopadhyay T~K and Datta A 2017 {\em the Journal of Physical Chemistry C\/}
  {\bf 121} 811--822

\bibitem{hernandez2010measurement}
Hernandez Y, Lotya M, Rickard D, Bergin S~D and Coleman J~N 2010 {\em
  Langmuir\/} {\bf 26} 3208--3213

\bibitem{plimpton1995fast}
Plimpton S 1995 {\em Journal of computational physics\/} {\bf 117} 1--19

\bibitem{lazar2013adsorption}
Lazar P, Karlicky F, Jurecka P, Kocman M, Otyepkova E, Safarova K and Otyepka M
  2013 {\em Journal of the American Chemical Society\/} {\bf 135} 6372--6377

\bibitem{patil2018adsorption}
Patil U and Caffrey N~M 2018 {\em The Journal of Chemical Physics\/} {\bf 149}
  094702

\bibitem{cheng1990computer}
Cheng A and Steele W 1990 {\em The Journal of chemical physics\/} {\bf 92}
  3858--3866

\bibitem{hardy2018design}
Hardy A, Dix J, Williams C~D, Siperstein F~R, Carbone P and Bock H 2018 {\em
  ACS nano\/} {\bf 12} 1043--1049

\bibitem{fu2013molecular}
Fu C and Yang X 2013 {\em Carbon\/} {\bf 55} 350--360

\bibitem{kamath2012silico}
Kamath G and Baker G~A 2012 {\em Physical Chemistry Chemical Physics\/} {\bf
  14} 7929--7933

\bibitem{jorgensen2005potential}
Jorgensen W~L and Tirado-Rives J 2005 {\em Proceedings of the National Academy
  of Sciences\/} {\bf 102} 6665--6670

\bibitem{dodda20171}
Dodda L~S, Vilseck J~Z, Tirado-Rives J and Jorgensen W~L 2017 {\em The Journal
  of Physical Chemistry B\/} {\bf 121} 3864--3870

\bibitem{dodda2017ligpargen}
Dodda L~S, Cabeza~de Vaca I, Tirado-Rives J and Jorgensen W~L 2017 {\em Nucleic
  acids research\/} {\bf 45} W331--W336

\bibitem{lorentz1881ueber}
Lorentz H 1881 {\em Annalen der physik\/} {\bf 248} 127--136

\bibitem{berthelot1898melange}
Berthelot D 1898 {\em Compt. Rendus\/} {\bf 126} 1703--1706

\bibitem{martinez2009packmol}
Mart{\'\i}nez L, Andrade R, Birgin E~G and Mart{\'\i}nez J~M 2009 {\em Journal
  of computational chemistry\/} {\bf 30} 2157--2164

\bibitem{ong2013python}
Ong S~P, Richards W~D, Jain A, Hautier G, Kocher M, Cholia S, Gunter D,
  Chevrier V~L, Persson K~A and Ceder G 2013 {\em Computational Materials
  Science\/} {\bf 68} 314--319

\bibitem{duarte2017approaches}
Duarte Ramos~Matos G, Kyu D~Y, Loeffler H~H, Chodera J~D, Shirts M~R and Mobley
  D~L 2017 {\em Journal of Chemical \& Engineering Data\/} {\bf 62} 1559--1569

\bibitem{mezei1987finite}
Mezei M 1987 {\em The Journal of chemical physics\/} {\bf 86} 7084--7088

\bibitem{kirkwood1935statistical}
Kirkwood J~G 1935 {\em The Journal of Chemical Physics\/} {\bf 3} 300--313

\bibitem{jorgensen1985monte}
Jorgensen W~L and Ravimohan C 1985 {\em The Journal of chemical physics\/} {\bf
  83} 3050--3054

\bibitem{liu2004calculation}
Liu P, Harder E and Berne B 2004 {\em The Journal of Physical Chemistry B\/}
  {\bf 108} 6595--6602

\bibitem{agosta2017diffusion}
Agosta L, Brandt E~G and Lyubartsev A~P 2017 {\em The Journal of chemical
  physics\/} {\bf 147} 024704

\bibitem{oyer2012stabilization}
Oyer A~J, Carrillo J~M~Y, Hire C~C, Schniepp H~C, Asandei A~D, Dobrynin A~V and
  Adamson D~H 2012 {\em Journal of the American Chemical Society\/} {\bf 134}
  5018--5021

\bibitem{mukhopadhyay2019exfoliation}
Mukhopadhyay T~K and Datta A 2019 {\em JOURNAL OF THE INDIAN CHEMICAL
  SOCIETY\/} {\bf 96} 753--766

\bibitem{frolov2012molecular}
Frolov A~I, Arif R~N, Kolar M, Romanova A~O, Fedorov M~V and Rozhin A~G 2012
  {\em Chemical science\/} {\bf 3} 541--548

\bibitem{terrones2016enhanced}
Terrones J, Kiley P~J and Elliott J~A 2016 {\em Scientific reports\/} {\bf 6}
  1--11

\bibitem{arunachalam2018graphene}
Arunachalam V and Vasudevan S 2018 {\em The Journal of Physical Chemistry C\/}
  {\bf 122} 1881--1888

\bibitem{gilliam2021evaluating}
Gilliam M~S, Yousaf A, Guo Y, Li D~O, Momenah A, Wang Q~H and Green A~A 2021
  {\em Langmuir\/} {\bf 37} 1194--1205

\bibitem{griffin2020effect}
Griffin A, Nisi K, Pepper J, Harvey A, Szyd{\l}owska B~M, Coleman J~N and
  Backes C 2020 {\em Chemistry of Materials\/} {\bf 32} 2852--2862

\bibitem{hu2021dispersant}
Hu C~X, Shin Y, Read O and Casiraghi C 2021 {\em Nanoscale\/} {\bf 13} 460--484

\bibitem{DDBONLINE}
Saturated liquid density
  \urlprefix\url{http://ddbonline.ddbst.de/DIPPR105DensityCalculation/DIPPR105CalculationCGI.exe?component=Acetone}

\bibitem{kim2019pubchem}
Kim S, Chen J, Cheng T, Gindulyte A, He J, He S, Li Q, Shoemaker B~A, Thiessen
  P~A, Yu B {\em et~al.\/} 2019 {\em Nucleic acids research\/} {\bf 47}
  D1102--D1109

\end{thebibliography}

\end{document}